# Effect of internal friction on transformation twin dynamics in $Sr_xBa_{1-x}SnO_3$ perovskite

Maren Daraktchiev,[*] Ekhard K. H. Salje, William T. Lee, and Simon A. T. Redfern

Department of Earth Sciences, University of Cambridge, Downing Street

Cambridge CB2 3EQ, United Kingdom

The dynamics of transformation twins in $Sr_xBa_{1-x}SnO_3$ (x=0.6,0.8) perovskite has been studied by dynamical mechanical analysis in three-point bend geometry. This material undergoes phase transitions from orthorhombic to tetragonal and cubic structures on heating. The mechanical loss signatures of the transformation twins include relaxation and frequency-independent peaks in the orthorhombic and tetragonal phases, with no observed energy dissipation in the cubic phase. The macroscopic shape, orientation and relative displacements of twin walls have been calculated from bending and anisotropy energies. The mechanical loss angle and distribution of relaxation time are discussed in term of bending modes of domain walls.

PACS numbers: 62.40.+i, 81.40.Jj, 91.60.-x

The attenuation of seismic waves at frequencies from 1 mHz to 1 Hz is expected to affect the geophysical signatures of mantle-forming minerals in the deep Earth, such as $(Mg,Fe)(Si,Al)O_3$ perovskites.[1-3] $(Mg,Fe)(Si,Al)O_3$ perovskite is orthorhombic at lower mantle pressures and temperatures[4], but is metastable at ambient T and P. Thus, structural analogues with a stable orthorhombic crystalline structure are needed for studying and modeling the low-frequency mechanical properties of perovskite. Ferroelastic perovskites, in particular those isostructural with $(Mg,Fe)(Si,Al)O_3$, develop mobile domain structures in their low-symmetry (ferroelastic) phases.[5] Mechanical loss or internal friction of these domain structures results in a dissipation peak related to the microstructure and its mobility. In the high symmetry (paraelastic) phase the ferroelastic transformation twins are absent and the dissipation of mechanical energy is (comparatively) negligible.[1, 5]

Earlier observations have stimulated much interest in the displacement of the domain walls on the mesoscopic scale, where the needle growth from crystal interfaces is a time – dependent self-adjustment towards a new equilibrium, dependent on temperature and applied external stress, $\sigma$. The dynamics of relaxation depend on a viscosity function $B$ [Pa.s/m] that originates from the internal structure of the domain wall and the domain wall-point defect interactions. As the viscous displacements of domain walls from their equilibrium positions are, by their very nature, dissipative in energy, they can be investigated by mechanical spectroscopy.[6]

Previous studies of domain wall trajectories have only considered the static case.[7] Here we extend that approach to include time dependence. The motion of a single transformation twin through an inhomogeneous concentration of point defects at given $T$ and $\sigma$ is described by the following time-dependent equation:

$$B\frac{dy(x,t)}{dt} = -S\frac{d^4y(x,t)}{dx^4} + U\frac{d^2y(x,t)}{dx^2} - \Pi y(x,t) + \sigma \qquad (1)$$

where $y(x,t)$ gives the trajectory of wall position in the elastically soft direction of the crystal, $B$ is the dragging force per unit length depending on the intrinsic viscosity of wall width, $\Pi$ is the Peierls force responsible for the lateral motion, and $S$ and $U$ are the bending and the anisotropy energies of wall segments, respectively.

Equation (1) determines the wall profile from the minimum condition that the total energy density[7]

$$E\big[y(x,t)\big] = \int \Big( E_{anisotropy} + E_{bending} + E_{Peierls} \Big).dx \qquad (2)$$

is given by the Euler-Lagrange equation

$$\frac{\delta E\big[y(x,t)\big]}{\delta y(x,t)} = F_{diss} + F_{ext} \qquad (3)$$

with dissipative $F_{diss} = Bdy(x,t)/dt$ and driving $F_{ext}$ forces per unit area of wall surface. $E_{bending} = S(d^2y(x,t)/dx^2)^2$ and $E_{anisotropy} = U(dy(x,t)/dx)^2$ measure the excess energies for bending and rotation of wall segments with respect to the elastically soft direction of the crystal. If dissipative forces act on a domain wall, then the wall dynamics are damped. In the approximation of defect-free bulk or weak domain wall-defect interactions, this damping is associated with the asymmetric growth of needles from the crystal interfaces due to the local structure of those interfaces and microscopic shearing of the sample. Here, we examine the effect of internal conformational arrangements of wall segments due to the movement of asymmetrical domain walls during an external stress perturbation.

In the steady-state limit ($Bdy/dt \rightarrow 0$), the dynamics of domain walls is suppressed in time by twin wall - defect interactions which, in general, determine needle trajectories and the symmetrical shapes of needle tips.[7] The tip remains rigidly pinned in the vicinity of defects for the time of any observation, but the trajectory may relax slowly causing a microscopic deformation of the sample at large distances from the pinning centre within a characteristic time greater than the lattice relaxation time. Such twin walls form, in particular, needles with shapes which do not depend on $T$, $P$ or the crystal structure in which they occur. Furthermore, low-frequency spectroscopy measurements and *in-situ* optical microscopy demonstrate that the needle motions towards equilibrium ($dy/dt \neq 0$) are subject to a Debye relaxation with a distribution of relaxation times.[1] The needles remain pinned on the crystal interfaces over a wide range of temperatures and stresses, while a comb of the symmetric and asymmetric needle tips is found to sweep between the sample boundaries yielding two mechanical responses in the Cole-Cole diagram: one with an anelastic signature described by a suppressed semicircle and other with a creep signature at low frequency which has yet to understand from a viewpoint of domain wall morphology and relaxation times.

$Sr_xBa_{1-x}SnO_3$ (x=0.6, 0.8) perovskite is a model system adopting the orthorhombic phase at room temperature. It undergoes a number of phase transitions to cubic symmetry with temperature (see Figure 1a). The low-symmetry orthorhombic (*Pnma, Imma*) and tetragonal (*I4/mcm*) structures are related to the ideal cubic ($Pm\overline{3}m$) structure of $Sr_xBa_{1-x}SnO_3$, via tilting of the $SnO_6$ octahedra about the cubic symmetry axes.[8] Experimentally, we have studied its mechanical response by dynamical mechanical analysis (DMA-7), which permits measuring the mechanical loss and elastic constants of $Sr_xBa_{1-x}SnO_3$ in parallel plate stress (PPS) and/or three-point bending (TPB) geometries.[3] As the elastic constants of perovskite are closely related to spontaneous strain induced during the $SnO_6$ titling, the mechanical loss spectrum in Figure 1 follows closely the phase transitions with temperature.[1, 9, 10] The peaks at 170 $^0$C ($Sr_{06}Ba_{04}SnO_3$) and 580 $^0$C ($Sr_{08}Ba_{02}SnO_3$) are spectroscopic features of the tetragonal-orthorhombic phase transitions (compare with the phase diagram on Figure 1a). The peak at 411 $^0$C for the 1 Hz mechanical loss spectrum of $Sr_{08}Ba_{02}SnO_3$ depends on frequency and is accompanied by a modulus anomaly. By using the peaks as markers of the frequency-temperature shifts, we estimated values of 1.85 eV and $10^{+14}$ s$^{-1}$ for the activation enthalpy $\Delta H_{act}$ and the attempt frequency $\omega_0$, respectively. These values are characteristics of a pinning-depinning transition which is mainly attributed to the interaction of domain walls with oxygen vacancies[11, 12] and cation interstitials[13] in perovskites. The apparent activation energy for those pinning sites decreases with $T$ in the ferroelastic phase as the energy landscape experienced by the domain walls becomes insensitive to the defect distribution with increasing temperature (wall width increases dramatically at $T_C$). Here, we explore whether needle growth is the origin for these apparent values of activation energy since needle displacements may contribute to the effective energy through spatial redistributions of bending modes due to the local structure of crystal surfaces.

Above 550 $^0$C ($Sr_{0.6}Ba_{0.4}SnO_3$) and 800 $^0$C ($Sr_{0.8}Ba_{0.2}SnO_3$), the measured loss angle is minimal, while the stiffness of material is maximal. Such a behavior is manifestation of a transition to the cubic symmetry in which the transformation twins not longer exist and domain wall dynamics have no bearing on dissipation.

On the macroscopic scale, the internal structure of domain walls is determined by the spatial variations of the order parameter below the phase transition.[5] The pulling and pushing of neighboring structural units at the boundary separating mesoscopic domains of different orientations (macroscopic spontaneous strains) induces displacement fields, which propagate in the elastically soft direction of the ferroelastic phase via a knock-on effect. In the approximation when no additional short-range interactions are considered, the coordinate of wall displacement $y(x,t)$ measured along the normal to domain wall is given by

$$y(x,t) = \int\limits_0^t \int\limits_0^l G(x,\xi,t-s)f(\xi)\delta(s)d\xi ds + \int\limits_0^t \int\limits_0^l G(x,\xi,t-s)\vartheta(\xi,s)d\xi ds \qquad (4)$$

In Eq. (4), $f(x)=y(x,0)$ accounts for the Cauchy initial conditions, $\delta(x)$ is the Dirac delta function, $G(x,\xi,t)$ is the Green's function and $\vartheta(\xi,s)$ is the non-homogeneous term in Eq. (1). The explicit solution of $G(x,\xi,t)$ is:[14]

$$G(x,\xi,t) = \frac{2}{l}\sum_n \sin\left(n\pi\frac{x}{l}\right)\sin\left(n\pi\frac{\xi}{l}\right)\exp\left(-\frac{\Pi}{B}t - \frac{U}{B}\left(\frac{n\pi}{l}\right)^2 t - \frac{S}{B}\left(\frac{n\pi}{l}\right)^4 t\right) \qquad (5)$$

The Green's function is calculated at $t{\geq}0$ for $0{\leq}x{\leq}l$, which accounts for the mesoscopic situations observed in perovskites[1, 3] when the twin boundaries propagate through the bulk under effect of applied stress while the surface of the crystal is allowed to relax so that $y=0$ at $x=0$ and $x=l$. It is relevant to define the relaxation time by $\tau=B/U(l/n\pi)^2=(1/n)^2\tau_U$ or $\tau=B/S(l/n\pi)^4=(1/n)^4\tau_S$, where $\tau_U$ and $\tau_S$ are the single Debye relaxation times, in the limiting cases that either the anisotropy or bending energy dominates the total energy. The Peierls contribution is not considered further in Eq. (5) as the domain walls mainly interact with defects at temperature at which $E_{Peierls} = \Pi.y(x,t)$ is much smaller than $E_{bending}$ and $E_{anisotropy}$. Thus, we neglect the influence of sample geometry at high temperatures ($E_{Peierls}{\rightarrow}0$) and calculate domain walls with universal shapes of needle tips. By contrast, at very low temperature the Peierls energy scales with the barrier height of the Landau double well potential and, thus dominates over all energies in the energy landscape. The wall trapping is determined by the lattice potential ($E_{Peierls}$) and/or point defects and the energy for thermal activation is calculated from the energy barrier in the Landau potential.[15]

The domain profiles at high temperatures are calculated for a stress field which is abruptly applied at $t=0$ and held to a constant value $\sigma_0$ at $t>0$. From Eqs. (4) and (5) one obtains:

$$y(x,t) = IT + \sigma^* \int\limits_0^t \int\limits_0^l G(x,\xi,s)d\xi.ds \qquad (6)$$

The initial trajectory (IT) determines the wall trajectory at $t=0$ for a stress level exceeding the threshold for nucleation. The profiles of domain walls from Eq. (6) are plotted in Figure 2. Simulations were carried out with $l\sim10$nm, $S/U/l^2\sim0.1$, $\sigma^*=\sigma_0/B\sim10^6$ nm.s$^{-1}$ and $\sigma_0\sim1000$ mN/(mm)$^2$, which are values typical for the needle profiles in ferroelastics.[5, 7] The parabolic tips of domain walls follow a Debye relaxation pattern towards an equilibrium position both whether the trajectories are dominated by the anisotropy (Figure 2a) or bending (Figure 2b) energies. The geometry of the needle tip is strongly dependent on the stress profile $\vartheta(\xi,s)$ applied on the perovskite surface. Figures 2a and 2b are examples of time-dependent wall evolution calculated from Eq. (6) for a homogeneous stress profile along the surface (the PPS mode). In TPB mode we induced an inhomogeneous stress profile through the sample by applying a dynamic force in the middle of Sr$_x$Ba$_{1-x}$SnO$_3$ surface, e.g. at $\xi_0=l/2$ precisely. Thus, by substituting $\vartheta(\xi,s)$d$\xi$d$s$ with $\sigma^*\delta(\xi-\xi_0)$d$\xi$d$s$ in Eq. (4) and we calculated the wall trajectory in the TPB geometry:

$$y(x,t) = IT + \sigma^* \int_0^t G(x,\xi_0,s)ds \qquad (7)$$

The results are plotted in Figures 2c and 2d for $E \approx E_{\text{anisotropy}}$ and $E \approx E_{\text{bending}}$, respectively. Time ranges from 2 to 100s were used for wall simulations in both geometries.

The wall trajectory in Figures 2c and 2d ends in a tip, which diverges from the parabolic one observed in Figures 2a and 2b. $E_{\text{anisotropy}}$ yields a straight trajectory around the needle tip for large $t$, which is identical to the steady-state profile calculated in previous works.[7] For small $t$, however, exponential needles develop in time with tip angles which vary with the needle half-width due to the inhomogeneous stress profile along $\xi$. They are not characteristic-length dependent as $E_{\text{Peierls}} \to 0$. They are due to the macroscopic sample deformation caused by the strain along the soft direction as a result of accommodation of wall relaxation at $\xi$=0 and $\xi$=l, as well as the absence of bending energy. Indeed, the excess of $E_{\text{bending}}$ does maximize the curvature of the wall trajectory at each point along $\xi$, as in Figure 2d, by superimposing a modified parabolic shape upon a straighter shape at the vicinity of the pinning centers on the surface. Strictly speaking, the maximum curvature of the parabolic profile at $\xi$=l/2 in Figure 2d is a result of the smooth connection of the rotated parts of the wall around the center.

In the PPS mode, neither exponential nor linear trajectories are calculated from Eq. (6). The profiles in Figure 2a are insensitive to the anisotropy energy since it is suppressed by the homogeneous sample deformation, which introduces parallel shift and constant curvature of domain walls at each point along the soft direction in the PPS mode. Furthermore, the wall displacements for $E_{\text{bending}}$ are also strengthened by this deformation making their trajectories in Figure 2b to look similar to ones calculated for $E_{\text{anisotropy}}$ in Figure 2a.

Building on the results in Figure 2 we observe that the wall displacement in the TPB mode depends on the magnitude of anisotropy/bending energy and the stress profile on the crystal surfaces. The boundary conditions applied in Eqs. (4) and (5) define a planar interface for needle growth at $t \geq 0$. In contrast, the surfaces of real experimental samples are highly rough and nonplanar. For the purposes of modeling we may consider a saw-toothed boundary along $\xi$ with an average slope, which depends on redistribution of normal stresses across the boundary plane.[16] The shear component of the stress is relaxed along $\xi$ (strong pinning sites on the surface) causing a displacement along the soft direction which is driven by the stress balance between the internal stresses acting on the boundary plane and $\sigma$. This results in a shift of the resulting stress from $\xi_0=l/2$ towards another point $\xi_0 \neq l/2$ which accounts for needles with tips at $\xi_0 \neq l/2$ and an asymmetric shape.

In Figure 3, the trajectories are calculated with a stress applied at $\xi_0=l/3$, which describes a model example of wall behaviors expected due to an inhomogeneous stress profile with $\sigma$ acting away from the center in the TPB geometry. The offset of the tip from the center is critical to the needle symmetry in crystals with a large $E_{\text{anisotropy}}$ and less critical in crystals with large $E_{\text{bending}}$. Indeed, for $E_{\text{bending}}$ the shift of the tip towards the center is balanced by the angular momentum for each length element that rotates the wall towards the center and the time-dependent sliding of each length element towards the equilibrium state, Figure 3b. Note that no tip shift with time is predicted for $E_{\text{anisotropy}}$ in Figure 3a because the wall dynamics are constrained along the elastically soft directions of the crystal, in particular, which is the axis of energy minimum of the conformational arrangements of wall segments.

The above results show that for non-planar interfaces the normal stresses cause a narrow distribution of the needle shapes for $E \approx E_{\text{bending}}$ (Figure 3b) around the wall trajectory of the most probable shape (e.g. one which is parabolic a tip in the center at equilibrium) and much broader one for $E \approx E_{\text{anisotropy}}$ (Figure 3a). Thus, a distribution of relaxation times around $\tau_m$ (the most probable value), together with broadening of the relaxation spectra around $\omega = 1/\tau_m$ in frequency space, is expected from the intrinsic effects of needle movements. Building on previous work on distribution of relaxation times in perovskites, we also know that dynamics associated with the needle motions throughout a non-uniform distribution of defects may generate a stretched exponential relaxation.[1] For the sake of completeness, an asymmetric profile in perovskites may be obtained by needle growth not only from surface defects, but also from bulk defects when needles split into pairs with smaller tip angles due to minimization of $E_{anisotropy}$ contribution.[5] The trapping of one of those tips into the stability field of point defects or lattice singularities (a probabilistic process) suppresses the tip displacement along the soft direction and promotes an asymmetric shape via lateral displacements of other tip into the bulk. It would be, however, difficult to distinguish needle dynamics due to surface and bulk defects quantitatively from mechanical spectroscopy alone, and additional observations, such as TEM would have to be made to confirm such a description.

For this reason, we chose here a qualitative description of the underling mechanism by deriving the probability density function of intrinsic movements of domain walls. The $E_{anisotropy}$ effects on the loss dynamics will be considered firstly. The complex compliance in $\omega$-space, $J^*(x,w)$, can be calculated in a straightforward manner[6] from

$$J^*(x,\omega) = J_U + i\omega \int\limits_0^\infty \left[ J(x,t) - J_U \right] \exp(-i\omega t) dt \qquad (8)$$

where $J(x,t) = J_U + y(x,t)/L/\sigma$. Thus, from Eq. (8) one gets

$$J^*(x,\omega) \equiv J_1 - iJ_2 = J_U + \frac{l^2}{LU}\sum_n \frac{\sin\left(n\pi\dfrac{x}{l}\right)}{n\pi}\frac{\sin\left(n\pi\dfrac{\xi_0}{l}\right)}{n\pi}\frac{1}{1+i\omega\tau_n} \qquad (9)$$

where $J_U = 1/E$ is the unrelaxed compliance and $L$ is the width of the sample in the elastically soft direction. Further, the discrete $\tau_n$ spectrum can be approximated by a continuous spectrum for time intervals closely spaced and relative to $1/\omega$. Examples of statistical developments with a continuous, rather than a discrete distribution of relaxation times are the motion of dislocations in the multi-well energy diagram dependent on the segregation of point defects on the dislocation line[17, 18] and the grain boundary sliding lubricated by an amorphous inter-granular layer causing stretched exponential relaxation due to the interaction between the relaxing species into the film[19]. Following Nowick and Berry[20] we define the distribution in terms of $\ln\tau$, rather than in $\tau$ itself. The expressions of $J_1$ and $J_2$ in the continuous limit are

$$\frac{J_1 - J_U}{\delta J} = \int_{-\infty}^{\infty} \frac{X\left(\ln\left(\tau/\tau_D\right)/\mu\right)}{1+\omega^2\tau^2}d\ln\left(\tau/\tau_D\right) \qquad (10a)$$

and

$$\frac{J_2}{\delta J} = \int_{-\infty}^{\infty} X\left(\ln\left(\tau/\tau_D\right)/\mu\right)\frac{\omega\tau}{1+\omega^2\tau^2}d\ln\left(\tau/\tau_D\right) \qquad (10b)$$

In Eq. (10), $\delta J = J_R - J_U = l^2/(LU)/2$ is the magnitude of the domain wall relaxation, $\mu$ is the distribution parameter proportional to the half-width of $X(\eta)$. Following Nowick and Berry[20] and Tobolsky[21], we substitute $\ln(\omega\tau_D)$ with z and $\ln(\tau/\tau_D) / \mu$ with $\eta$:

$$\frac{J_1 - J_U}{\delta J} = \int_{-\infty}^{\infty} \frac{X(\eta)}{1 + \exp(2z + 2\mu\eta)} d\eta \equiv f_1(z, \mu) \tag{11a}$$

and

$$\frac{J_2}{\delta J} = \int_{-\infty}^{\infty} \frac{X(\eta)}{\cosh(z + \mu\eta)} d\eta \equiv f_2(z, \mu) \tag{11b}$$

Equations (11a) and (11b) are similar to those used by Nowick and Berry[20] and Tobolsky[21] for the box and lognormal distributions of relaxation times. The normalized probability density function of domain walls

$$X\left(\frac{\ln(\tau/\tau_D)}{\mu}\right) \equiv \frac{\left(\sin\left(\pi\frac{\ln(\tau/\tau_D)}{\mu}\right)/\pi\ln(\tau/\tau_D)\right)^2}{\int_{-\infty}^{\infty}\left(\sin\left(\pi\frac{v}{\mu}\right)/\pi v\right)^2 dv} = \frac{1}{\mu}\left(\frac{\sin\left(\pi\frac{\ln(\tau/\tau_D)}{\mu}\right)}{\pi\frac{\ln(\tau/\tau_D)}{\mu}}\right)^2 \tag{12}$$

is a corollary of the Fourier transforms of the integral kernel (Eq. (5)). It assigns the distribution density of relaxation time around some mean value $\tau_U$ in the interval $[\tau, \tau + d\tau]$ as a result of the probability density of finding a needle trajectory defined by $\tau$ around the most probable needle bending mode defined by $\tau_U$ in the range between $\tau$ and $\tau+d\tau$. The distribution parameter was introduced in the $sinc^2$ function (Eq.12) through the offset of needle tips from the center (Figure 3), which follows from the redistribution of the normal stress along $\xi$ as a consequence of surface inhomogeneity.

In Figure 4, the relaxation peaks in $Sr_{08}Ba_{02}SnO_3$ well-modeled by a distribution of Debye peaks $tan\phi(z,\mu)$ calculated from Eq. (11) in terms of variable $1/T$, rather than in $ln(\omega\tau_D)$ itself, for thermally activated atomic motions. We calculated $\mu$=10.3 and $\delta J/J_U$=0.35 from the relative peak width, $r_2(\mu)\equiv\Delta T^{-1}(\mu)/\Delta T^{-1}(0)$=3.7 and the relative peak height, $2f_2(0,\mu) \equiv J_2(0)/\delta J$=0.28, where $\Delta T^{-1}(0)$ and $J_2(0)$ are the half-width and the relaxation strength of the single Debye peak. For the fit we used a Debye peak with relative peak width given by $\Delta H_{act}/r_2(\mu)$, as well as the values of $(tan\phi)_P$ and $T_P$ at the maximum of the peak. The internal friction peak in Figure 4a and therefore the relaxation process are completely characterized by $\omega_0$, $\Delta H_{act}$, $\mu$ and $\delta J/J_U$.

The values of $\delta J/J_U$, $f_2(0,\mu)$ and $r_2(\mu)$ calculated from Eq. (11) differ from ones calculated and tabulated by Nowick for the lognormal distribution.[20] This difference is a corollary of the maxima existing at $x = \pm3/2\pi$ around the main maximum at $x$=0 in $sinc^2(x)$ which results in an additional statistical broadening of relaxation times for large $\mu$. For small $\mu$, however, the broadening of the domain walls is described by the Gaussian-like distribution[20] as the central maximum in $sinc^2(x)$ smears out the effect of other maxima. The transition from small values of $\mu$ to larger ones gives a transition in $J_2$-$J_1$ space which does not result from the onset of low-frequency creep in the material but does result from the different quantitative description of the wall dynamics in terms of distributions of relaxation times. Experimental studies on superelasticity show that $\mu$ changes with temperature[1], e.g. it decreases with increasing $T$ (the distribution of the relaxation time is more likely to be in the activation energy rather than in the limit relaxation time.[1,20]), through a number of factors (wall width, wall density, surface roughness) which are intrinsic function of temperature.[3, 10] Thus, the systematic dependence of $\mu$ with $T$ may lead to a transition in the $J_2$-$J_1$ diagram[1], as well as positive shift in the effective value of the activation energy.

In conclusion, for $E \approx E_{\text{anisotropy}}$ the needle trajectory of ferroelastic domain walls gives rise to a distribution of relaxation times for wall dynamics around $\tau_U$. This, however, is not found when $E \approx E_{\text{bending}}$. It follows from (4) and (5) that the $E_{\text{bending}}$ effects can be introduced in $J_2$ and $J_1$ spectra by replacing $X(\eta)$ with $X(\eta)/\eta^2$ into Eq. (11). For integral operations it is more convenient to normalize latter and write in terms of $\delta(\eta) \equiv \lim_{\varepsilon \to 0} X(\eta)\varepsilon/(\eta^2+\varepsilon)$. Both integrals reduce to the Debye equations for $E \approx E_{\text{bending}}$. Hence, needle dynamics are governed by a single Debye relaxation time $\tau_S$, which determines a needle trajectory of a modified parabolic shape. No broadening of relaxation peaks in terms of relaxation times exists in a crystal with $E \approx E_{\text{bending}}$. Additionally, it follows from $S/U \sim l^2$ for $l \sim 100$ nm and $\tau_S = 2\pi/\omega_S = B/Sl^4/\pi^4$ that the maximum in the energy dissipation is described by a relaxation peak in the kHz-frequency range, which is much higher than the frequency range in Figures 1 and 4, and the frequency of seismic waves, although the influence of pressure in the deep Earth on this relation has yet to be determined.

This work was financially supported by NERC, grant number NER/A/S/2003/00537.

*Electronic address: mdar04@esc.cam.ac.uk

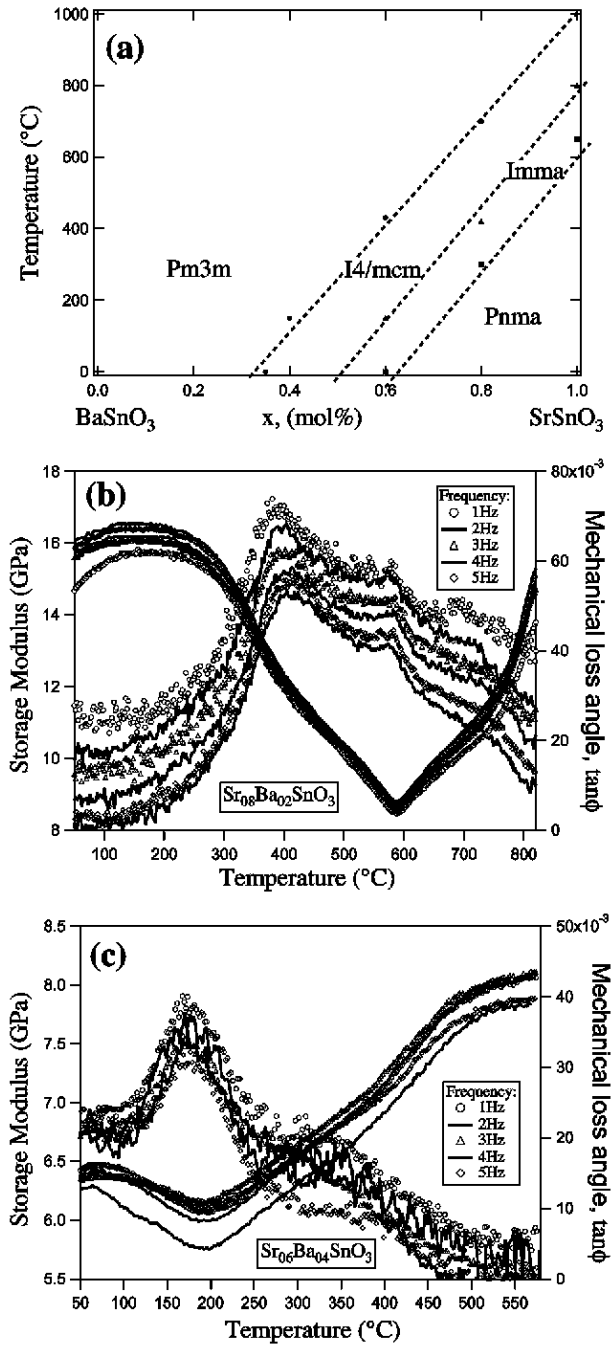

FIG 1. (a) Temperature-composition phase diagram for the solid-solution $Sr_xBa_{1-x}SnO_3$ stannate; Isochronal mechanical loss spectra of (b) $Sr_{06}Ba_{04}SnO_3$ and (c) $Sr_{08}Ba_{02}SnO_3$ at frequency between 1 and 6 Hz. For the cubic phase above (b) 800 °C and (c) 550 °C the modulus saturates and the loss drops to a negligible value, so the data are not reported.

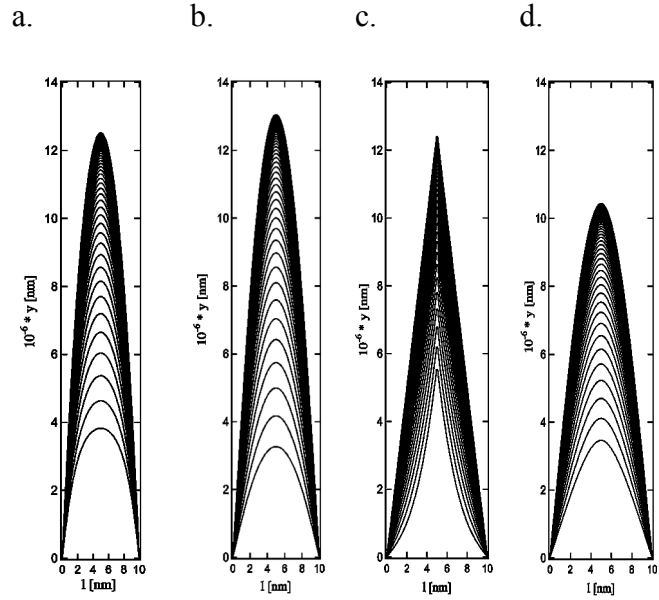

FIG 2. Needle profiles calculated for $t$ from 0 to 100 s in 2 s steps. Each trajectory corresponds to a different time. The stress is abruptly applied at $\xi_0$ ($t$=0) and held to a constant value $\sigma_0$ at $t > 0$. (a) PPS mode, anisotropy dominates over bending energy; (b) PPS mode, bending dominates over anisotropy energy; (c) and (d) Similar concepts of energies but in TPB mode

a.

b.

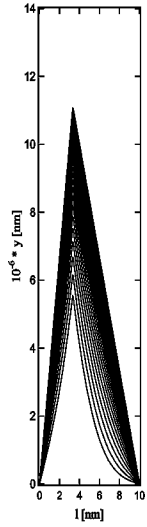
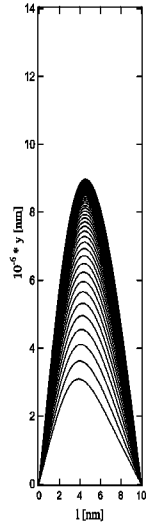

FIG 3. Needle profile for a stress at $\xi_0 = l/3$ in TPB mode. (a) Anisotropy energy dominates bending's; (b) Bending energy is larger than anisotropy's.

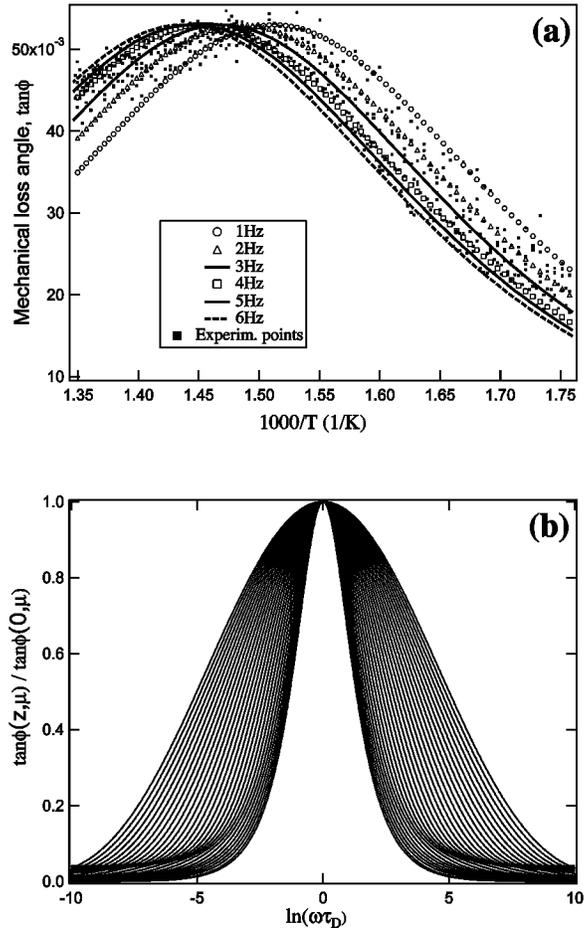

FIG 4. (a) Mechanical loss peaks in $Sr_{08}Ba_{02}SnO_3$ in the domain freezing $T$ are fitted with distributed Debye peaks. (b) Normalized distributed Debye peaks $tan\phi(z, \mu)/tan\phi(0, \mu)$ calculated for $\mu$=0-6. A single Debye relaxation process corresponds to $\mu$=0